\documentclass{aa}
\usepackage{graphicx} 
\usepackage[colorlinks=true,
            linkcolor=blue,
            citecolor=blue,
            urlcolor=blue,
            anchorcolor=blue,
            linkcolor=blue]{hyperref}

\usepackage{amsmath,amsfonts,amssymb,mathtools}
\usepackage{booktabs}
\usepackage{multirow}
\usepackage[dvipsnames]{xcolor}
\usepackage{soul}
\usepackage{orcidlink}

\newcommand{\fnl}{f_{\rm NL}}

\begin{document}

\title{Hierarchical summaries for primordial non-Gaussianities}

\author{M.~S.~Cagliari\thanks{\email{marina.cagliari@lapth.cnrs.fr}}\inst{\ref{inst1}}\,\orcidlink{0000-0002-2912-9233}\and A. Bairagi\inst{\ref{inst2}}\,\orcidlink{0009-0009-3089-052X}\and
B.~Wandelt\inst{\ref{inst3},\ref{inst4},\ref{inst2},\ref{inst5}}\,\orcidlink{0000-0002-5854-8269}}

\authorrunning{M.~S.~Cagliari et al.}

\institute{Laboratoire d’Annecy de Physique Theorique (LAPTh), CNRS/USMB, 99 Chemin de Bellevue BP110 - Annecy - F-74941 - ANNECY CEDEX - FRANCE\label{inst1}
\and 
CNRS \& Sorbonne Universit\'{e}, Institut d’Astrophysique de Paris (IAP), UMR 7095, 98 bis bd Arago, F-75014 Paris, France\label{inst2}
\and
Department of Physics and Astronomy, Johns Hopkins University, 3400 North Charles Street, Baltimore, MD 21218, USA\label{inst3}
\and
Department of Applied Mathematics and Statistics, Johns Hopkins University, 3400 North Charles Street, Baltimore, MD 21218, USA\label{inst4}
\and
Center for Computational Astrophysics, Flatiron Institute, 162 5th Avenue, New York, NY 10010, USA\label{inst5}
}

\date{}

\abstract{The advent of Stage IV galaxy redshift surveys such as DESI and \textit{Euclid} marks the beginning of an era of precision cosmology, with one key objective being the detection of primordial non-Gaussianities (PNG), potential signatures of inflationary physics. In particular, constraining the amplitude of local-type PNG, parameterised by $\fnl$, with $\sigma_{\fnl} \sim 1$, would provide a critical test of single versus multi-field inflation scenarios. While current large-scale structure and cosmic microwave background analyses have achieved $\sigma_{\fnl} \sim 5$--$9$, further improvements demand novel data compression strategies. We propose a hybrid estimator that hierarchically combines standard $2$-point and $3$-point statistics with a field-level {neural summary}, motivated by recent theoretical work that shows that such a combination is nearly optimal, disentangling primordial from late-time non-Gaussianity. We employ \textsc{PatchNet}, a convolutional neural network that extracts small-scale information from sub-volumes (patches) of the halo number density field while large-scale information is retained via the power spectrum and bispectrum. Using \textsc{Quijote-PNG} simulations, we evaluate the Fisher information of this combined estimator across various redshifts, halo mass cuts, and scale cuts. Our results demonstrate that the inclusion of patch-based field-level compression always enhances constraints on $\fnl$, reaching gains of $30$--$45\%$ at low $k_{\rm max}$ ($\sim 0.1 \, h \, \text{Mpc}^{-1}$),  and capturing information beyond the bispectrum. This approach offers a computationally efficient and scalable pathway to tighten the PNG constraints from forthcoming survey data.}


\maketitle

\section{Introduction} \label{sec:intro}

With the results and observations from stage IV surveys, such as the data release 1 of the Dark Energy Spectroscopic Instrument \citep[DESI;][]{desi, DESI:2024hhd} and the first observations by \textit{Euclid} \citep{Euclid:2024yrr}, we have entered the era of precision large-scale structure (LSS) cosmology. The key objective of the new generation surveys, such as DESI and \textit{Euclid}, but also SPHEREx \citep{spherex}, and the Vera C. Rubin Legacy Survey of Space and Time \citep[LSST;][]{lsst}, is to stress test the standard $\Lambda$CDM cosmological model, both at late times, measuring the dark energy equation of state; and at early times, detecting primordial features that would be smoking guns of inflation. In particular, inflationary models predict small deviations from Gaussianity in the primordial density distribution, called primordial non-Gaussianities (PNG), the footprint of which we should be able to observe in the late-time distribution of galaxies. 

Different inflation models produce different types of PNG, see \citet{Baumann:2009ds} for a review. In this work, we are interested in the PNG of local type, which we parametrise with $\fnl$. This parameter controls the amplitude of the non-linear relation between a mean-zero Gaussian field, $\varphi$, and the primordial gravitational potential, $\Phi_P = \varphi + f_{\rm NL} (\varphi^2-\langle\varphi^2\rangle)$. We know that local PNG should be very close to zero in the case of single-field inflation \citep{Maldacena:2002vr,Creminelli:2004yq,Cabass:2016cgp}, while $\fnl \sim \mathcal{O}(1)$ for multi-field inflation models \citep{Senatore:2010wk,Alvarez:2014}. Therefore, a detection or non-detection of local PNG with $\sigma_{\fnl} \sim 1$ would rule out single-field inflation or constraint multi-field models, respectively. For this reason, intensive efforts took place in the cosmological community to measure and improve our constraining power over this observable using different probes. 

Currently, the most stringent constraints for $\fnl$ come from the measurements of the $3$-point function of the cosmic microwave background (CMB) by \textit{Planck} and are consistent with zero with $\sigma_{\fnl} \sim 5$ \citep{2020A&A...641A...9P}. We expect the new generation of CMB surveys to tighten these bounds by a factor of $2$, but not to reach $\sigma_{\fnl} \sim 1$ \citep{CMB-S4:2016ple}. On the other hand, local PNG affect the late-time LSS both at the $2$- and $3$-point function level, which, in Fourier space, correspond to the power spectrum and the bispectrum \citep{Dalal2008,Slosar2008,Desjaques2010}. The most stringent bounds from LSS measurements come from the power spectrum analysis of the first data release of DESI and reads $\sigma_{\fnl} \sim 9$ \citep{DESI:2024oco}, which corresponds to a $40\%$ improvement over the previous power spectrum-based measurement that used the quasar sample of data release 16 of the extended Baryon Oscillation Spectroscopic Survey \citep{Cagliari:2023mkq} and over $50\%$ improvement with respect to previous measurements based on power spectrum and bispectrum \citep[e.g.,][]{DAmico:2022gki,Cabass:2022ymb}. This shows a first glimpse of the constraining power of stage IV surveys. These results will improve even more when the complete surveys become available.

At the same time as gathering new data, we may attempt to improve our constraints on $\fnl$, or other cosmological parameters, by devising new estimators that retain more information than the standard approach. For example, the authors of ~\citet{Castorina2019} computed the optimal quadratic estimator to measure the cosmological $\fnl$ signal from the LSS power spectrum multipoles. Additionally, many works combine the power spectrum and the bispectrum \citep{Cagliari:2025rqe} and utilise effective field theory \citep[EFT; see][for a review]{Cabass:2022avo} to extend the scale range we can probe with these statistics to smaller scales. Alternatively, we could use field-level algorithms that bypass the data compression into $n$-point statistics. These methods are based either on forward modelling \citep{Andrews:2022nvv,Andrews2024} or machine learning algorithms applied to the late-time galaxy distribution \citep{Kvasiuk:2024gbz}, also reconstructing the initial condition \citep{Chen2025}, or to the CMB \citep{Nagarajappa2024}. Several detailed simulation studies have explored the choice (or combinations) of summary statistics to extract PNG information \citep{Coulton:2022rir,Jung:2022gfa,Coulton:2023ouk,Jung:2024esv} and possibly guide the community effort.

The work of \citet{Giri2023} produced highly informative estimates of $\fnl$ from the non-linear matter field by training a neural network to compute local estimates of $\sigma_8$ and modelling auto- and cross spectra of the $\sigma_8$-field and the matter. The methodology relies on a bias model specifically for local non-Gaussianity, which neatly bypasses the need for running training simulations with non-Gaussian initial conditions but it is not obvious how to generalise this approach to other forms of primordial non-Gaussianity.

In this work, we propose a method that approximates global field-based analysis by hierarchically combining standard $n$-point statistics with a local, neural summary that extracts field-level information. The main idea is that we can use the power spectrum and bispectrum optimally to extract large-scale information, and machine learning based data compression to extract small-scale information in the non-linear regime. Given a large cosmological volume, we expect that on large scales the majority of the $\fnl$ information will be encoded in the $2$- and $3$-point functions; on the other hand, on small scales, a field-level algorithm should extract more information than the power spectrum and bispectrum alone \citep{Bairagi:2025ytq}. Aside from the application to non-Gaussianity, we go beyond the dark matter field and focus on catalogues of halos. Neither of them are directly observable, but the halo field is a better proxy than the dark matter field for the signal that is achievable with observations.

This hierarchical combination of global statistics with local field-level analysis is theoretically well-motivated for the problem of disentangling non-linear evolution from primordial non-Gaussianity. Recent work demonstrated that locality fundamentally protects primordial signals from contamination by late-time gravitational nonlinearities \citep{Baumann_2022}. Specifically, primordial non-Gaussianity creates genuinely non-local correlations between widely separated points (generated during inflation), while late-time gravitational evolution is constrained by causality to produce only local effects that cannot mimic these long-range correlations.

This locality principle has important implications for data analysis. Fisher analysis shows that higher-order bias parameters become increasingly orthogonal to primordial signals in the perturbative regime; and ``only quadratic nonlinearities affect the map-level analysis, while all higher-order nonlinearities decouple'' \citep{Baumann_2022}. This suggests that combining 2-point and 3-point statistics with local field-level information naturally incorporates the constraints from higher-order correlations without requiring their explicit computation. The map-level approach thus captures the essential information content on primordial non-Gaussianity more efficiently than computing progressively higher-order correlation functions.

In this spirit, our algorithm measures the power spectrum and bispectrum of the whole observational volume, while it runs a field-level analysis on patches. We analyse the patches with a convolutional neural network \citep[CNN;][]{LeCun:89}, named \textsc{PatchNet} \citep{Bairagi:2025ytq}, which summarises the field-level information into one number and then aggregates over all the patches to get a single-value compression. Then, the final summary statistic is the combination of the $2$-point and $3$-point summaries with the field-level compression. This combination was shown in \citep{Bairagi:2025ytq} to produce state-of-the-art information extraction for cosmological parameters in the non-linear regime.

On a practical level, this approach has the advantages of a machine learning field-level analysis without the shortcoming of being too computationally expensive: the network acts only on small sub-volumes of the full computational volume that we dub patches, and it greatly reduces the number of simulations to train the networks compared with a full field-based analysis, since dividing the volume into patches greatly increases the training set size.

In this paper, we estimate the $\fnl$ information content of this new estimator applied to the \textsc{Quijote-PNG} dark matter halos \citep{Coulton:2022qbc} in a cube of $1 \, \text{Gpc}^3\, h^{-3}$. We perform a detailed study of the Fisher information of combining the power spectrum with local patches and the combined power spectrum, bispectrum, and patches as a function of redshift, halo mass cut, and scale cut for the standard summary statistics. 

The main finding of this work is that, consistent with theoretical expectations, patch-based data compression consistently provides additional information compared to standard summary statistics. The most significant improvements occur at the lowest $k_{\rm max}$ for the $n$-point summary statistics, with gains ranging between $30\%$ and $45\%$. Furthermore, when comparing the power spectrum and bispectrum constraints to those derived from patch-combined estimators, the patches consistently enhance the results, demonstrating they capture information beyond the $3$-point function. 

The paper is organised as follows: in Sect.~\ref{sec:data} we present the simulated data we employ for the analysis; in Sect.~\ref{sec:methods} we introduce the Fisher formalism, the $n$-point statistics we use and describe the architecture of \textsc{PatchNet}. We discuss our results in Sect.~\ref{sec:results} and conclude in Sect.~\ref{sec:conclusions}.

\section{Simulations} \label{sec:data}

The \textsc{Quijote} simulations \citep{Villaescusa-Navarro:2019bje} are a set of N-body simulations run with \texttt{GADGET-III}, which was originally developed for the Aquarius project \citep{2008MNRAS.391.1685S} and is an improved version of the \texttt{GADGET-II} code \citep{gadget2}. They are specifically designed to test machine learning applications to cosmological data. They start from initial conditions at $z=127$ generated with the \texttt{2LPTIC} code\footnote{\url{https://cosmo.nyu.edu/roman/2LPT/}.} and simulate five redshift snapshots ($z=0, 0.5, 1, 2,$ and $3$) with a comoving volume of $1000^3 \, \text{Mpc}^3 \, h^{-3}$ and a fiducial resolution of $512^3$ particles.\footnote{High-resolution $1024^3$-particle and low-resolution $128^3$-particle simulations are available.} The main data products of the \textsc{Quijote} simulations are the particle snapshots over which two halo finders were run. For this work, we used the halo catalogues produced with the Friend-of-Friend \citep[FoF;][]{FoF} halo finder.

The original \textsc{Quijote} dataset spans over a range of $\Lambda$CDM and $\nu w$CDM cosmologies. Recently, a new set of simulations has become available, the \textsc{Quijote-PNG} \citep{Coulton:2022qbc} simulations, which cover four different types of PNG: LSS orthogonal, CMB orthogonal, equilateral, and local PNG. The initial conditions of the \textsc{Quijote-PNG} simulation were generated with the \texttt{2LPTPNG} code \citep{Coulton:2022qbc}.\footnote{\url{https://github.com/dsjamieson/2LPTPNG}.} For this work, we only used the datasets related to local PNG. For the training of \textsc{PatchNet}, we used the $1000$ $\fnl$ local Latin hypercube simulations. In this set $\fnl$ uniformly spans the interval $(-300,300)$. Then, we use $5000$ realisations of the fiducial cosmology for the Fisher information study, where $\fnl=0$. Finally, we also used the two sets of $500$ simulations with $\fnl$ displaced from the fiducial value. These simulations have $\fnl=-50$ or $\fnl=50$ and are used to estimate the derivative of the summary statistics as a function of $\fnl$. For all these simulations, except for $\fnl$ all the other cosmological parameters are fixed to the fiducial values, $\Omega_m = 0.3175$, $\Omega_b = 0.049$, $h = 0.6711$, $n_s = 0.9624$, $\sigma_8 = 0.834$, $w = -1$, and $M_{\nu} = 0 \, \text{eV}$.

Considering the redshift range probed by current surveys \citep[e.g., the Euclid Wide Survey;][]{EuclidWide,Euclid:2024yrr}, we are mainly interested in the results at $z=1$, which should better represent the clustering observed by this experiment. In addition, we studied the redshift dependence of the algorithm performance at the other four redshifts available in the Quijote dataset, $z=0, \, 0.5, \, 2$, and $3$. We tested two scenarios: first, the case where the same mass cut is applied at all redshifts, namely, we selected halos with $M_{\text{h}} > 3.12 \times 10^{13} \, M_{\odot} \, h^{-1}$, and the alternative case where the number density is constant at different redshifts. In this last case, we chose at every redshift a halo mass cut, $M_{\text{halo}} > M_{\text{min}}$, that would fix the number density of the halos in the fiducial cosmology to the number density at $z=3$ ($n(z) \sim 5 \times 10^{-6} \, h^3 \, \text{Mpc}^{-3}$), which is the least populated snapshot. We present the threshold masses, $M_{\text{min}}$, in Table \ref{tab:Mmin}.

\begin{table}
    \centering
    \caption{Threshold masses at different redshifts.}
    \begin{tabular}{cc} \toprule 
         Redshift  & $M_{\text{min}} \, [M_{\odot} \, h^{-1}]$ \\ \midrule
         $3.0$ & $0.0$  \\
         $2.0$ & $3.47 \times 10^{13}$ \\ 
         $1.0$ & $1.05 \times 10^{14}$ \\
         $0.5$ & $1.91 \times 10^{14}$ \\
         $0.0$ & $3.47 \times 10^{14}$ \\ \bottomrule
    \end{tabular}
    \vspace{1ex}

    {\raggedright \small \textbf{Notes.} The values of $M_{\text{min}}(z)$ keep the number density constant ($n(z) \sim 5 \times 10^{-6} \, h^3 \, \text{Mpc}^{-3}$) at the different redshifts in the fiducial cosmology.  \par}
    \label{tab:Mmin}
\end{table}

\section{Methods} \label{sec:methods}

\subsection{Fisher Information} \label{sec:fisher}

It is possible to evaluate the amount of information and the constraining power of an estimator through the Fisher information analysis. The Fisher information matrix is defined as follows,
\begin{equation}
    F_{ij} = \left( \frac{\partial \mathbf{x}}{\partial \theta_i} \right)^T \mathbf{C}^{-1} \left( \frac{\partial \mathbf{x}}{\partial \theta_j} \right) \, ,
    \label{eq:fisher}
\end{equation}
where $\mathbf{x}$ is a summary statistics of the optimal unbiased estimator of the parameters $\theta_i$ and $\theta_j$, and $\mathbf{C}^{-1}$ is the inverse of the covariance matrix of the summary statistics $\textbf{x}$. The expected variance of parameter $i$ is $\sigma^2_i = (F^{-1})_{ii}$. Therefore, to estimate the constraining power of a given summary statistic, we first need to compute the covariance matrix of the estimator and its derivatives with respect to the parameters of interest, which in this study is only $\fnl$. We estimated both $\mathbf{C}^{-1}$ and $\frac{\partial \mathbf{x}}{\partial \fnl}$ numerically. We used $5000$ \textsc{Quijote} simulations in the fiducial cosmology to estimate the covariance matrix, $\mathbf{\hat{C}}^{-1}$, which we also correct with the Hartlap factor \citep{Hartlap-factor},
\begin{equation}
    \mathbf{C}^{-1} = \frac{n_{\rm s} - n_{\rm x} - 2}{n_{\rm s} - 1} \, \mathbf{\hat{C}}^{-1} \, ,
    \label{eq:hartlap}
\end{equation}
to account for the non-infinite number of data we use to estimate the covariance matrix. In Eq.~(\ref{eq:hartlap}), $n_{\rm s}$ is the number of simulations used to compute the covariance and $n_{\rm x}$ is the length of the summary statistic.
Finally, we estimate the derivatives numerically from the finite difference,
\begin{equation}
    \frac{\partial \mathbf{x}}{\partial \fnl} = \frac{\bar{\mathbf{x}}\left( \fnl^{\text{fid}} + \delta\fnl \right) - \bar{\mathbf{x}}\left( 
    \fnl^{\text{fid}} - \delta\fnl \right)}{2 \, \delta\fnl} \, ,
    \label{eq:deriv}
\end{equation}
where, $\bar{\mathbf{x}}\left( \fnl^{\text{fid}} \pm \delta\fnl \right)$ are the mean summary statistics of the simulations with $\fnl = \pm 50$. Then, as $\fnl^{\text{fid}} = 0$, $\delta\fnl = 50$.\footnote{To test convergence, we run the power spectrum Fisher analysis varying the number of simulations used to compute the numerical derivatives. We verified that the results are stable from 300 simulations.}

\subsection{Summary statistics: Power spectrum and Bispectrum} \label{sec:pkbk}

\begin{figure*}
    \centering
    \includegraphics[width=0.49\linewidth]{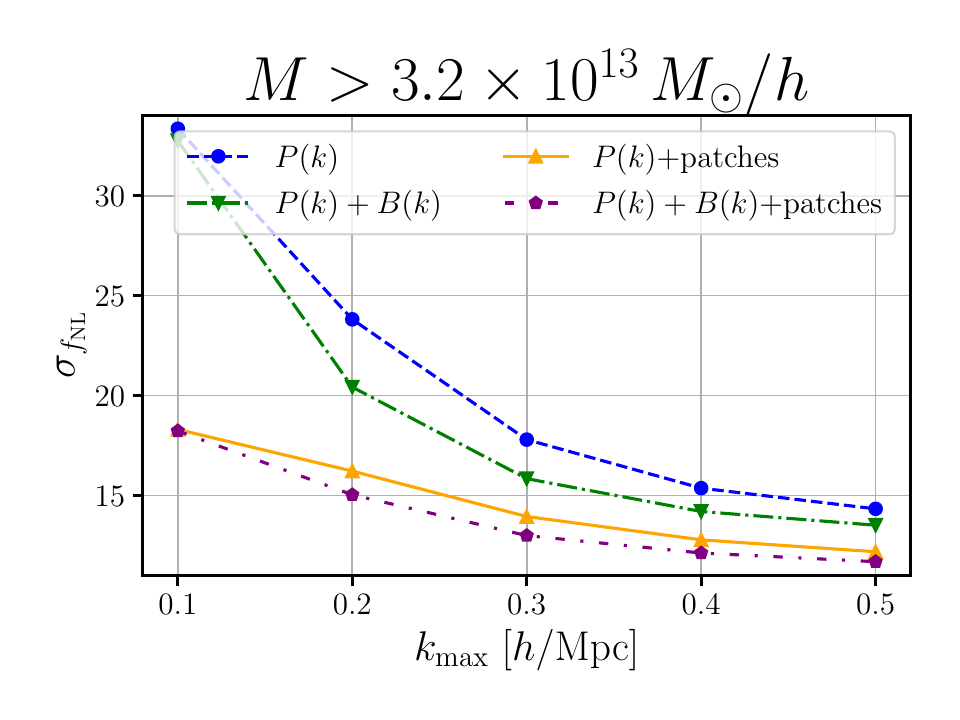}\includegraphics[width=0.49\textwidth]{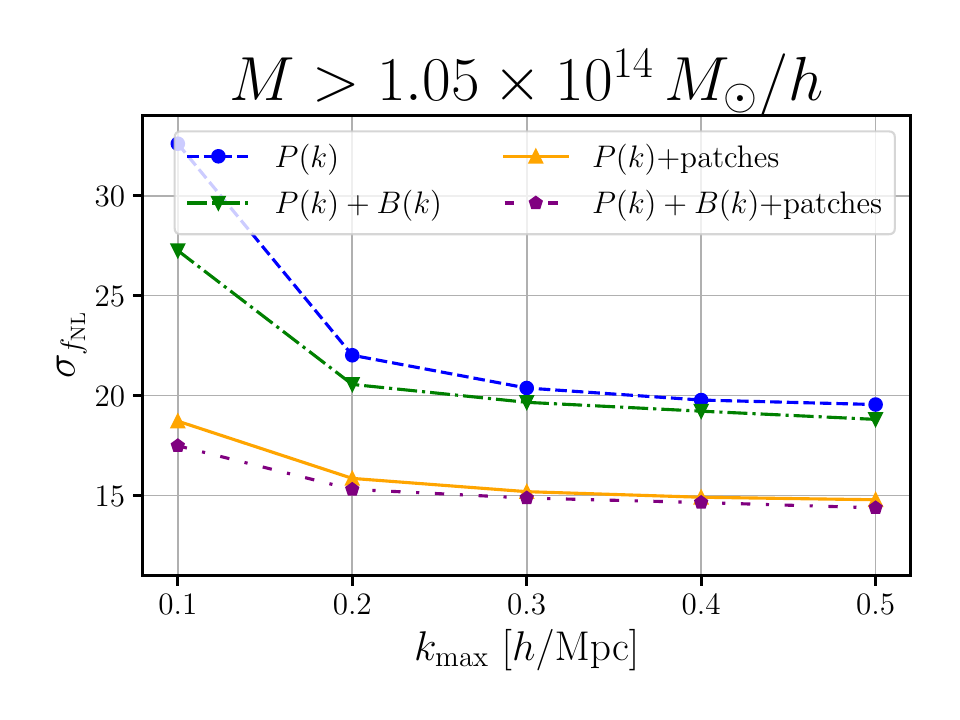}
    \caption{Expected uncertainty for the different summary statistics at $z=1$ for two halo-mass cuts as a function of the power spectrum and bispectrum $k_{\text{max}}$. In both cases, the \textsc{PatchNet} information (patches) improves the constraining power of the power spectrum and the power spectrum combined with bispectrum analyses.}
    \label{fig:res-z1}
\end{figure*}
In this work, we aim to combine large- and small-scale information to improve the constraining power over local $\fnl$ measurements. A first solution is to use effective field theory to increase the minimum scale we can probe with standard summary statistics. The alternative approach we explored is to use field-level data compression based on machine learning to extract small-scale information, while the large-scale information still comes from standard summary statistics.

The summary statistics we utilise are the power spectrum and the bispectrum. Given the halo over-density, $\delta_{\rm h}$, the power spectrum corresponds to the $2$-point correlation function in Fourier space, and we define it as follows
\begin{equation}
    \langle \delta_{\rm h}(\mathbf{k}) \, \delta_{\rm h}^*(\mathbf{k'}) \rangle = (2 \, \pi)^3 \, \delta^{\rm D}(\mathbf{k} - \mathbf{k'}) \, P(k) \, ,
    \label{eq:Pk-def}
\end{equation}
where $\delta^{\rm D}$ is the $3$-dimensional Dirac's delta function. Similarly, we define the halo bispectrum, which is the $3$-point correlation function in Fourier space, as
\begin{equation}
    \langle \delta_{\rm h}(\mathbf{k}_1) \, \delta_{\rm h}(\mathbf{k}_2) \, \delta_{\rm h}(\mathbf{k}_3) \rangle = (2 \, \pi)^3 \, \delta^{\rm D}(\mathbf{k}_1 + \mathbf{k}_2 + \mathbf{k}_3) \, B(\mathbf{k}_1, \mathbf{k}_2, \mathbf{k}_3) \, .
    \label{eq:Bk-def}
\end{equation}

To measure the power spectra and bispectra of the simulated boxes, we used the public codes \texttt{Pylians3} \citep{Pylians} and \texttt{pySpectrum} \citep{2015PhRvD..92h3532S,2020JCAP...03..040H}, respectively. We measure the power spectra on a $512^3$ grid from the fundamental wave number of the box, $k_{\rm f} \sim 0.006 \, h \, \text{Mpc}^{-1}$, to $k_{\text{max}} \sim 0.8 \, h \, \text{Mpc}^{-1}$ with a $k_{\rm f}$ linear binning; on the other hand, we use a $256^3$ grid for the bispectrum, which we measure with a $3 \, k_{\rm f}$ linear binning from $k_{\rm min} = 3 \, k_{\rm f}$ to $k_{\text{max}} \sim 0.5 \, h \, \text{Mpc}^{-1}$.

\subsection{Field level compression with \textsc{PatchNet}} \label{sec:patchnet}

We employ field-level data compression, using a CNN to extract small-scale information following the approach of \citet{Bairagi:2025ytq}.  The network does not analyse the whole simulated box; it only analyses a sub-cube, which we dub `patch' and call the network \textsc{PatchNet}. The \textsc{PatchNet} approach has two main advantages: first, it reduces the memory requirement for the CNN while maintaining the pixel (or better voxel) resolution, as it processes a smaller volume. Second, the number of training data greatly increases as we can cut out a set of $N_{\text{sb}}$ patches from one simulated box to train the network, effectively increasing the amount of training data by a factor of $N_{\text{sb}}$. As we will explain below, we train \textsc{PatchNet} to compress each patch into one informative statistic, namely the value of $\fnl$, and then aggregate these statistics by averaging across the $N_{\text{sb}}$ patches of a given simulated box. We then concatenated this aggregated information with the power spectrum or the combination of the power spectrum and bispectrum.

\textsc{PatchNet} has the following architecture: first, there are three blocks composed of a convolutional layer and a three-dimensional average pooling layer, then their output is flattened and processed by three dense layers before the final output. The convolutional layers, as well as the average pooling, have a $3 \times 3 \times 3$ kernel with a stride of $1$ and $0$ padding. The first convolutional layer has $8$ filter in output, and the filters are increased by a factor of $4$ in the following layers. After the flattening, the latent features are compressed into $512$ neurons and subsequently halved in number by the other dense layers. Both the convolutional and dense layers have a LeakyReLU \citep{Maas2013RectifierNI} with a negative slope of $0.5$ as activation function. We use a modified version of the mean squared error as loss function, which reads as follows
\begin{equation}
    \text{Loss}(\mathbf{y}, \mathbf{\hat{y}}) = \sum_i^{n_{\rm b}} \frac{1}{n_{\rm b}} \left( \sum_j^{n_{\text{sb}}} \frac{1}{n_{\text{sb}}} y_{j,i} - \sum_j^{n_{\text{sb}}} \frac{1}{n_{\text{sb}}} \hat{y}_{j,i} \right)^2 \, ,
    \label{eq:loss}
\end{equation}
where $\mathbf{y}$ are the outputs of \textsc{PatchNet}, $\mathbf{\hat{y}}$ are the labels ($\fnl$), $n_{\rm b}$ is the batch size, i.e., the number of simulations we load, and $n_{\text{sb}}$ is the number of patches we use for each simulation. We use $n_{\rm b}=32$ and $n_{\text{sb}} = 64$; therefore, a batch actually contains $n_{\rm b} \times n_{\text{sb}}$ different inputs. 

\begin{figure*}
    \centering
    \includegraphics[width=.98\linewidth]{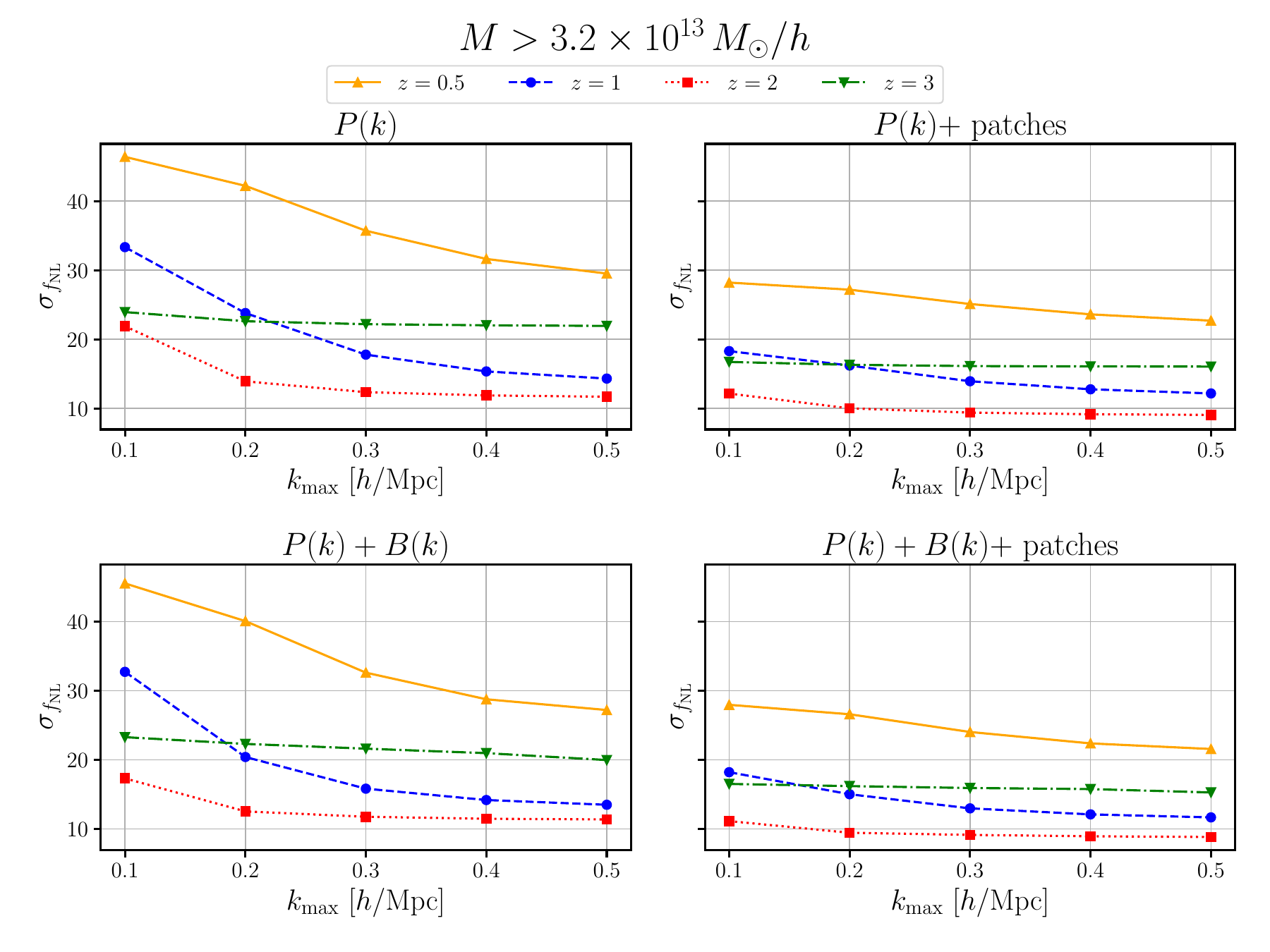}
    \caption{Expected uncertainty for the different summary statistics as a function of the power spectrum and bispectrum $k_{\text{max}}$ at different redshift snapshots. At all redshifts, we applied the same halo-mass cut, $M_{\text{h}} > 3.2 \times M_{\odot} \, h^{-1}$. The patch information adds constraining power at every redshift.}
    \label{fig:Mcut-zdep}
\end{figure*}
Before inputting the halo catalogues to the network, we preprocess them. First, we divide the $1000$ simulations with varying $\fnl$ into the training, validation, and test sets, which we select with a standard split of $75\%$, $15\%$, and $10\%$ of the data, respectively. Then, we build the patches from the simulated halo catalogue density contrast, which we compute by painting the halos on a $128^3$ grid with the cloud in cell (CIC) algorithm. Note that during the painting process, we do not use any halo mass information. We divide the $3$-dimensional number density contrast into $16^3$ non-overlapping cubes.\footnote{This choice is conservative and produces an underestimation of the information content. Indeed, the authors of \citet{Bairagi:2025ytq} showed that the use of overlapping patches increases information by $15\%$ to $20\%$.} This results in $512$ patches for each simulation that have a physical dimension of $125^3 \, \text{Mpc}^3 \, h^{-3}$ and a resolution of $8 \, \text{Mpc} \, h^{-1}$, which in Fourier space correspond to $k_{\text{min}} \sim 0.05 \, h \, \text{Mpc}^{-1}$ and $k_{\text{max}} \sim 0.4 \, h \, \text{Mpc}^{-1}$. Finally, before feeding the density contrast patches to the network, we centre and scale them with the mean and standard deviations computed from the density contrast of the fiducial cosmology realisations. We note that we computed these two quantities for each redshift and mass cut configuration (see Sect.~\ref{sec:data}).

\section{Results} \label{sec:results}

\begin{figure*}
    \centering
    \includegraphics[width=.98\linewidth]{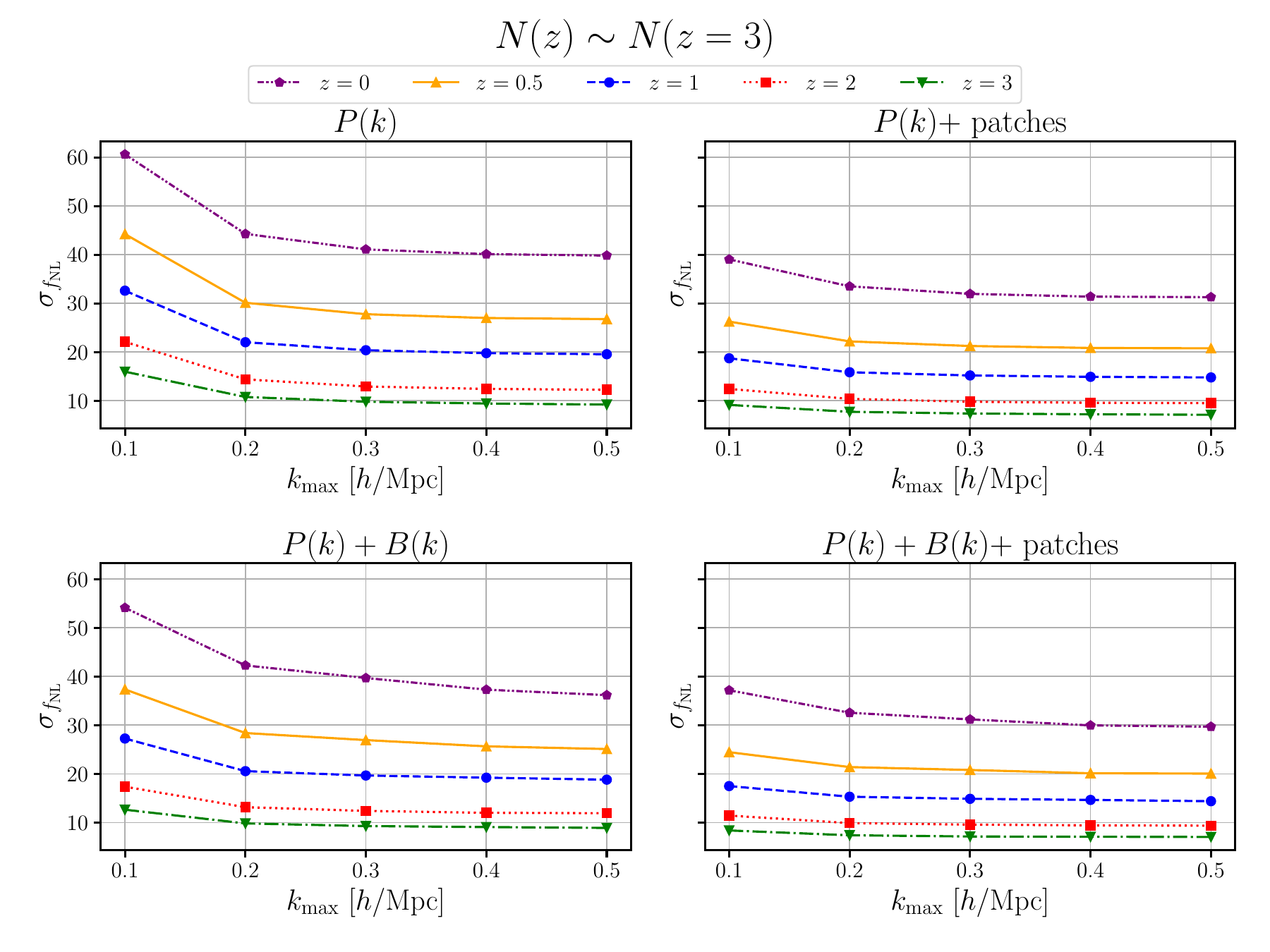}
    \caption{Expected uncertainty for the different summary statistics at fixed number density over redshifts as a function of the power spectrum and bispectrum $k_{\text{max}}$. At each redshift, we selected a halo-mass cut that has in the fiducial cosmology the same number density of $z=3$ (see Table~\ref{tab:Mmin}). As in Fig.~\ref{fig:Mcut-zdep}, the patch information improves the constraining power of the standard summary statistics.}
    \label{fig:nconst-zdep}
\end{figure*}
In this section, we present and discuss the results of the Fisher information analysis. In particular, we compare the information content of the power spectrum and bispectrum with their combination with the patch information. 

Figure \ref{fig:res-z1} shows the results of the Fisher information analysis at redshift $z=1$ for two mass cuts as a function of the $k_{\text{max}}$ of the power spectrum and bispectrum. Adding patch information always produces tighter bounds than the standard summary statistics alone. Moreover, we also observe an improvement in the combination of the power spectrum and the patches (solid orange line with triangular markers) with respect to the combined power spectrum and bispectrum (dash-dotted green line with reverse triangular markers), showing that \textsc{PatchNet} extracts information not captured by the 2-point and 3-point functions. This is also backed up by the fact that even though the improvement due to the patch information decreases when $k_{\rm max}$ increases, as the minimum scale probed by the standard summary statistics gets closer to the minimum scale of the patches ($k_{\rm max}^{\rm patch} \sim 0.4 \, h \, \text{Mpc}^{-1}$), we still observe an improvement when $k_{\rm max}^{\rm patch} < k_{\rm max}^{P,B}$ point again in the direction that the patches use information beyond the $3$-point function. Nevertheless, the bispectrum still brings a small amount of information that the network cannot extract, as the combination of the three statistics (sparse dash-dotted purple line with pentagonal markers) is always slightly lower than the combination of the power spectrum and patches alone. This behaviour is not unexpected as the bispectrum has access to the very large-scale squeezed triangle configuration, which contains non-Gaussian information that the patches cannot access due to their limited size. 

Comparing the left and right panels we see that the overall information content of the smaller mass cut ($M_{\text{min}} = 3.2 \times 10^{13} \, M_{\odot} \, h^{-1}$, left panel) is higher than the larger mass cut ($M_{\text{min}} = 1.05 \times 10^{14} \, M_{\odot} \, h^{-1}$, right panel). This is because when the mass cut is smaller, the halo field has lower shot noise and contains more information. However, the improvement related to the patch information is greater in the case of the larger mass cut for $k_{\text{max}} \geq 0.2 \, h \, \text{Mpc}^{-1}$. This means that, in terms of real objects, this method is more effective for highly biased objects, e.g., quasars, which we usually utilise to produce surveys with the large volumes that are required for $\fnl$ measurements.

We present the constraining power of the different statistics as a function of redshift in Figs. \ref{fig:Mcut-zdep} and \ref{fig:nconst-zdep}. In Fig. \ref{fig:Mcut-zdep} we kept the $M_{\text{min}}$ constant, while in Fig. \ref{fig:nconst-zdep} we varied it as a function of redshift to keep the number density of the catalogues equal to the halo number density at $z=3$. The feature that is common to both plots is that the addition of the patch information always makes the expected error at $k_{\text{max}} = 0.1 \, h \, \text{Mpc}^{-1}$ comparable to, if not better than, the constraint of the standard summary statistics alone at $k_{\text{max}} = 0.5 \, h \, \text{Mpc}^{-1}$. This is an interesting result, considering that modelling the power spectrum and bispectrum becomes more involved as we enter the non-linear regime, $k_{\text{max}} > 0.1 \, h \, \text{Mpc}^{-1}$, requiring a large number of simulations, or EFT to model the observed data. The use of patch information can greatly reduce the effort in the modelling of the standard summary statistics, while requiring a smaller simulation budget than neural estimators acting on the full field.

Finally, the authors of \citet{Jung:2023kjh} have shown that there is a high information content related to $\fnl$ in the halo mass function. As some halo mass information is retained in the density contrast, \textsc{PatchNet} may be just learning the relation between the halo mass function and $\fnl$, making the field-level analysis superfluous. To check if this is the case, we trained \textsc{PatchNet} keeping the number of objects in the catalogue fixed. The difference between this analysis and the one with constant density is that in the latter, the mass cut was chosen at the fiducial cosmology (with $\fnl=0$) and used for all the other simulated cosmologies. Therefore, there was a variation in the number density as a function of $\fnl$. On the other hand, for this test, we keep the number of objects fixed even when we vary the value of $\fnl$. Doing so, we remove the halo mass function information. Even in this configuration, \textsc{PatchNet} learns a relationship between the density contrast field and $\fnl$. In this case, the Pearson correlation coefficient is $r_{\text{w/o HMF}} \sim 0.78$, which we compare with $r_{\text{w HMF}} \sim 0.99$ for the standard configuration. This shows that the network extracts information beyond the halo mass function, as $r_{\text{w HMF}} > r_{\text{w/o HMF}}$. Nevertheless, the halo mass function remains highly informative, with $r \gg 0.50$.

\section{Conclusions} \label{sec:conclusions}

In this work, we propose an alternative approach to EFT up to large $k_{\rm max}$ or more standard ML-based field-level algorithms to analyse galaxy redshift surveys. The algorithm combines the large-scale information encoded in the $n$-point summary statistics, such as the power spectrum or bispectrum, with the small-scale information obtained by averaging over the compressed field-level information of small patches of the simulated box, the \textsc{PatchNet} approach \citep{Bairagi:2025ytq}. The main advantage of this approach is that the analysis of small patches is computationally lighter than analysing the whole box; additionally, by cutting the simulated boxes into patches, we greatly increase the number of training examples for the \textsc{PatchNet} \citep{Bairagi:20250313755B}. We estimated the Fisher information of this combined summary statistics to extract local PNG information. 

We tested the algorithm on the FoF halo catalogues of the \textsc{Quijote-PNG} simulations. We compared its performance with the more standard analyses based on the power spectrum or the combination of the power spectrum and bispectrum. We performed this comparison for different values of $k_{\rm max}$ for the $n$-point statistics and studied it as a function of redshift and for different halo mass cuts.

Our preliminary study shows that the patch-based data compression always provides additional information relative to the standard summary statistics. We observe the largest improvements when the power spectrum and the bispectrum are truncated at  $k_{\rm max} = 0.1 \, h \, \text{Mpc}^{-1}$, where the small-scale structure in the patches provides the most complementary information. At this scale, depending on the redshift, mass cut, and the statistics combination, the improvement varies between $30\%$ and $45\%$. Moreover, when we compare the power spectrum plus bispectrum bounds with the patch-combined bounds (both power spectrum with patches and power spectrum plus bispectrum with patches), again the patches always bring an improvement, proving that they contain information beyond the $3$-point function. Nevertheless, we also observed that combining the bispectrum with the \textsc{PatchNet} compression also provides an improvement, which shows that the large-scale bispectrum encodes information that the patches miss due to their limited size.

A first extension to this work would be to test our method's performance for other forms of PNG  and the effect of marginalising over cosmological parameters. Second, to fully understand the potential of the patch-based approach, it would be of interest to compare its information content with other alternative summary statistics, e.g., wavelet scattering transform \citep[e.g.,][]{Peron2024} or a marked power spectrum \citep[e.g.,][]{Marinucci2025}.
Additionally, to ready this approach for application to data, we will need to introduce a galaxy model in the simulations and understand how to treat the lightcone, the survey mask, and systematic effects.  We plan to study these issues in future work.

\begin{acknowledgements}
    We thank Francisco Villaescusa-Navarro, William Coulton, and Michele Liguori for their feedback on the first version of this manuscript. MSC thanks `Fondazione Angelo della Riccia' for financial support. The work of MSC is supported by the Agence Nationale de la Recherche (ANR) grant n. ANR-23-CPJ1-0160-01. AB acknowledges support from the Simons Foundation as part of the Simons Collaboration on Learning the Universe. The Flatiron Institute is supported by the Simons Foundation. All neural training and post-analysis have been done on IAP’s Infinity cluster. The authors thank St\'ephane Rouberol for his efficient management of this facility.
\end{acknowledgements}

\bibliographystyle{aa_url}
\bibliography{main}

\begin{thebibliography}{51}
\expandafter\ifx\csname natexlab\endcsname\relax\def\natexlab#1{#1}\fi

\bibitem[{{Abazajian} {et~al.}(2016){Abazajian}, {Adshead}, {Ahmed}, {Allen}, {Alonso}, {Arnold}, {Baccigalupi}, {Bartlett}, {Battaglia}, {Benson}, {Bischoff}, {Borrill}, {Buza}, {Calabrese}, {Caldwell}, {Carlstrom}, {Chang}, {Crawford}, {Cyr-Racine}, {De Bernardis}, {de Haan}, {di Serego Alighieri}, {Dunkley}, {Dvorkin}, {Errard}, {Fabbian}, {Feeney}, {Ferraro}, {Filippini}, {Flauger}, {Fuller}, {Gluscevic}, {Green}, {Grin}, {Grohs}, {Henning}, {Hill}, {Hlozek}, {Holder}, {Holzapfel}, {Hu}, {Huffenberger}, {Keskitalo}, {Knox}, {Kosowsky}, {Kovac}, {Kovetz}, {Kuo}, {Kusaka}, {Le Jeune}, {Lee}, {Lilley}, {Loverde}, {Madhavacheril}, {Mantz}, {Marsh}, {McMahon}, {Meerburg}, {Meyers}, {Miller}, {Munoz}, {Nguyen}, {Niemack}, {Peloso}, {Peloton}, {Pogosian}, {Pryke}, {Raveri}, {Reichardt}, {Rocha}, {Rotti}, {Schaan}, {Schmittfull}, {Scott}, {Sehgal}, {Shandera}, {Sherwin}, {Smith}, {Sorbo}, {Starkman}, {Story}, {van Engelen}, {Vieira}, {Watson}, {Whitehorn}, \& {Kimmy Wu}}]{CMB-S4:2016ple}
{Abazajian}, K.~N., {Adshead}, P., {Ahmed}, Z., {et~al.} \href{https://ui.adsabs.harvard.edu/abs/2016arXiv161002743A}{2016, arXiv:1610.02743}

\bibitem[{{Adame} {et~al.}(2025){Adame}, {Aguilar}, {Ahlen}, {Alam}, {Alexander}, {Allende Prieto}, {Alvarez}, {Alves}, {Anand}, {Andrade}, {Armengaud}, {Avila}, {Aviles}, {Awan}, {Bahr-Kalus}, {Bailey}, {Baltay}, {Bault}, {Behera}, {BenZvi}, {Beutler}, {Bianchi}, {Blake}, {Blum}, {Bonici}, {Brieden}, {Brodzeller}, {Brooks}, {Buckley-Geer}, {Burtin}, {Calderon}, {Canning}, {Carnero Rosell}, {Cereskaite}, {Cervantes-Cota}, {Chabanier}, {Chaussidon}, {Chaves-Montero}, {Chebat}, {Chen}, {Chen}, {Claybaugh}, {Cole}, {Cuceu}, {Davis}, {Dawson}, {de la Macorra}, {de Mattia}, {Deiosso}, {Dey}, {Dey}, {Ding}, {Doel}, {Edelstein}, {Eftekharzadeh}, {Eisenstein}, {Elbers}, {Elliott}, {Fagrelius}, {Fanning}, {Ferraro}, {Ereza}, {Findlay}, {Flaugher}, {Font-Ribera}, {Forero-S{\'a}nchez}, {Forero-Romero}, {Frenk}, {Garcia-Quintero}, {Garrison}, {Gazta{\~n}aga}, {Gil-Mar{\'\i}n}, {Gontcho}, {Gonzalez-Morales}, {Gonzalez-Perez}, {Gordon}, {Green}, {Gruen}, {Gsponer}, {Gutierrez}, {Guy}, {Hadzhiyska}, {Hahn}, {Hanif},
  {Herrera-Alcantar}, {Honscheid}, {Howlett}, {Huterer}, {Ir{\v{s}}i{\v{c}}}, {Ishak}, {Joyce}, {Juneau}, {Kara{\c{c}}ayl{\i}}, {Kehoe}, {Kent}, {Kirkby}, {Kong}, {Koposov}, {Kremin}, {Krolewski}, {Lahav}, {Lai}, {Lan}, {Landriau}, {Lang}, {Lasker}, {Le Goff}, {Le Guillou}, {Leauthaud}, {Levi}, {Li}, {Lodha}, {Magneville}, {Manera}, {Margala}, {Martini}, {Matthewson}, {Maus}, {McDonald}, {Medina-Varela}, {Meisner}, {Mena-Fern{\'a}ndez}, {Miquel}, {Moon}, {Moore}, {Moustakas}, {Mudur}, {Mueller}, {Mu{\~n}oz-Guti{\'e}rrez}, {Myers}, {Nadathur}, {Napolitano}, {Neveux}, {Newman}, {Nguyen}, {Nie}, {Niz}, {Noriega}, {Padmanabhan}, {Paillas}, {Palanque-Delabrouille}, {Pan}, {Penmetsa}, {Percival}, {Pieri}, {Pinon}, {Poppett}, {Porredon}, {Prada}, {P{\'e}rez-Fern{\'a}ndez}, {P{\'e}rez-R{\`a}fols}, {Rabinowitz}, {Raichoor}, {Ram{\'\i}rez-P{\'e}rez}, {Ramirez-Solano}, {Rashkovetskyi}, {Ravoux}, {Rezaie}, {Rich}, {Rocher}, {Rockosi}, {Roe}, {Rosado-Marin}, {Ross}, {Rossi}, {Ruggeri}, {Ruhlmann-Kleider}, {Samushia},
  {Sanchez}, {Saulder}, {Schlafly}, {Schlegel}, {Schubnell}, {Seo}, {Shafieloo}, {Sharples}, {Silber}, {Slosar}, {Smith}, {Sprayberry}, {Tan}, {Tarl{\'e}}, {Taylor}, {Trusov}, {Vaisakh}, {Valcin}, {Valdes}, {Valogiannis}, {Vargas-Maga{\~n}a}, {Verde}, {Walther}, {Wang}, {Wang}, {Weaver}, {Weaverdyck}, {Wechsler}, {Weinberg}, {White}, {Wilson}, \& {Yi}}]{DESI:2024hhd}
{Adame}, A.~G., {Aguilar}, J., {Ahlen}, S., {et~al.} 2025, \href{http://dx.doi.org/10.1088/1475-7516/2025/07/028}{\textcolor{blue}JCAP}, \href{https://ui.adsabs.harvard.edu/abs/2025JCAP...07..028A}{2025, 028}

\bibitem[{{Alvarez} {et~al.}(2014){Alvarez}, {Baldauf}, {Bond}, {Dalal}, {de Putter}, {Dor{\'e}}, {Green}, {Hirata}, {Huang}, {Huterer}, {Jeong}, {Johnson}, {Krause}, {Loverde}, {Meyers}, {Meerburg}, {Senatore}, {Shandera}, {Silverstein}, {Slosar}, {Smith}, {Zaldarriaga}, {Assassi}, {Braden}, {Hajian}, {Kobayashi}, {Stein}, \& {van Engelen}}]{Alvarez:2014}
{Alvarez}, M., {Baldauf}, T., {Bond}, J.~R., {et~al.} \href{https://ui.adsabs.harvard.edu/abs/2014arXiv1412.4671A}{2014, arXiv:1412.4671}

\bibitem[{{Andrews} {et~al.}(2024){Andrews}, {Jasche}, {Lavaux}, {Leclercq}, {Finelli}, {Akrami}, {Ballardini}, {Karagiannis}, {Valiviita}, {Bartolo}, {Ca{\~n}as-Herrera}, {Casas}, {Granett}, {Pace}, {Paoletti}, {Porqueres}, {Sakr}, {Sapone}, {Aghanim}, {Amara}, {Andreon}, {Baccigalupi}, {Baldi}, {Bardelli}, {Bonino}, {Branchini}, {Brescia}, {Brinchmann}, {Camera}, {Capobianco}, {Carbone}, {Carretero}, {Castellano}, {Castignani}, {Cavuoti}, {Cimatti}, {Colodro-Conde}, {Congedo}, {Conselice}, {Conversi}, {Copin}, {Courbin}, {Courtois}, {Da Silva}, {Degaudenzi}, {De Lucia}, {Di Giorgio}, {Dinis}, {Dubath}, {Duncan}, {Dupac}, {Dusini}, {Farina}, {Farrens}, {Faustini}, {Ferriol}, {Frailis}, {Franceschi}, {Galeotta}, {Gillis}, {Giocoli}, {G{\'o}mez-Alvarez}, {Grazian}, {Grupp}, {Haugan}, {Holmes}, {Hormuth}, {Hornstrup}, {Hudelot}, {Ili{\'c}}, {Jahnke}, {Jhabvala}, {Joachimi}, {Keih{\"a}nen}, {Kermiche}, {Kiessling}, {Kubik}, {Kunz}, {Kurki-Suonio}, {Ligori}, {Lilje}, {Lindholm}, {Lloro}, {Maiorano}, {Mansutti},
  {Marggraf}, {Markovic}, {Martinelli}, {Martinet}, {Marulli}, {Massey}, {Medinaceli}, {Mei}, {Mellier}, {Meneghetti}, {Merlin}, {Meylan}, {Moresco}, {Moscardini}, {Neissner}, {Niemi}, {Nightingale}, {Padilla}, {Paltani}, {Pasian}, {Pedersen}, {Pettorino}, {Pires}, {Polenta}, {Poncet}, {Popa}, {Pozzetti}, {Raison}, {Rebolo}, {Renzi}, {Rhodes}, {Riccio}, {Romelli}, {Roncarelli}, {Saglia}, {S{\'a}nchez}, {Sartoris}, {Schirmer}, {Schneider}, {Schrabback}, {Secroun}, {Sefusatti}, {Serrano}, {Sirignano}, {Sirri}, {Stanco}, {Steinwagner}, {Tallada-Cresp{\'\i}}, {Taylor}, {Tereno}, {Toledo-Moreo}, {Torradeflot}, {Tutusaus}, {Valenziano}, {Vassallo}, {Verdoes Kleijn}, {Veropalumbo}, {Wang}, {Weller}, {Zamorani}, {Zucca}, {Burigana}, {Scottez}, {Spurio Mancini}, \& {Viel}}]{Andrews2024}
{Andrews}, A., {Jasche}, J., {Lavaux}, G., {et~al.} \href{https://ui.adsabs.harvard.edu/abs/2024arXiv241211945A}{2024, arXiv:2412.11945}

\bibitem[{{Andrews} {et~al.}(2023){Andrews}, {Jasche}, {Lavaux}, \& {Schmidt}}]{Andrews:2022nvv}
{Andrews}, A., {Jasche}, J., {Lavaux}, G., \& {Schmidt}, F. 2023, \href{http://dx.doi.org/10.1093/mnras/stad432}{\textcolor{blue}\mnras}, \href{https://ui.adsabs.harvard.edu/abs/2023MNRAS.520.5746A}{520, 5746}

\bibitem[{{Bairagi} \& {Wandelt}(2025)}]{Bairagi:2025ytq}
{Bairagi}, A. \& {Wandelt}, B. \href{https://ui.adsabs.harvard.edu/abs/2025arXiv250903165B}{2025, arXiv:2509.03165}

\bibitem[{{Bairagi} {et~al.}(2025){Bairagi}, {Wandelt}, \& {Villaescusa-Navarro}}]{Bairagi:20250313755B}
{Bairagi}, A., {Wandelt}, B., \& {Villaescusa-Navarro}, F. \href{https://ui.adsabs.harvard.edu/abs/2025arXiv250313755B}{2025, arXiv:2503.13755}

\bibitem[{Baumann(2011)}]{Baumann:2009ds}
Baumann, D. 2011, in Theoretical Advanced Study Institute in Elementary Particle Physics: Physics of the Large and the Small, \href{}{523--686}

\bibitem[{{Baumann} \& {Green}(2022)}]{Baumann_2022}
{Baumann}, D. \& {Green}, D. 2022, \href{http://dx.doi.org/10.1088/1475-7516/2022/08/061}{\textcolor{blue}JCAP}, \href{https://ui.adsabs.harvard.edu/abs/2022JCAP...08..061B}{2022, 061}

\bibitem[{{Cabass} {et~al.}(2023){Cabass}, {Ivanov}, {Lewandowski}, {Mirbabayi}, \& {Simonovi{\'c}}}]{Cabass:2022avo}
{Cabass}, G., {Ivanov}, M.~M., {Lewandowski}, M., {Mirbabayi}, M., \& {Simonovi{\'c}}, M. 2023, \href{http://dx.doi.org/10.1016/j.dark.2023.101193}{\textcolor{blue}Physics of the Dark Universe}, \href{https://ui.adsabs.harvard.edu/abs/2023PDU....4001193C}{40, 101193}

\bibitem[{{Cabass} {et~al.}(2022){Cabass}, {Ivanov}, {Philcox}, {Simonovi{\'c}}, \& {Zaldarriaga}}]{Cabass:2022ymb}
{Cabass}, G., {Ivanov}, M.~M., {Philcox}, O. H.~E., {Simonovi{\'c}}, M., \& {Zaldarriaga}, M. 2022, \href{http://dx.doi.org/10.1103/PhysRevD.106.043506}{\textcolor{blue}\prd}, \href{https://ui.adsabs.harvard.edu/abs/2022PhRvD.106d3506C}{106, 043506}

\bibitem[{{Cabass} {et~al.}(2017){Cabass}, {Pajer}, \& {Schmidt}}]{Cabass:2016cgp}
{Cabass}, G., {Pajer}, E., \& {Schmidt}, F. 2017, \href{http://dx.doi.org/10.1088/1475-7516/2017/01/003}{\textcolor{blue}JCAP}, \href{https://ui.adsabs.harvard.edu/abs/2017JCAP...01..003C}{2017, 003}

\bibitem[{{Cagliari} {et~al.}(2025){Cagliari}, {Barberi-Squarotti}, {Pardede}, {Castorina}, \& {D'Amico}}]{Cagliari:2025rqe}
{Cagliari}, M.~S., {Barberi-Squarotti}, M., {Pardede}, K., {Castorina}, E., \& {D'Amico}, G. 2025, \href{http://dx.doi.org/10.1088/1475-7516/2025/07/043}{\textcolor{blue}JCAP}, \href{https://ui.adsabs.harvard.edu/abs/2025JCAP...07..043C}{2025, 043}

\bibitem[{{Cagliari} {et~al.}(2024){Cagliari}, {Castorina}, {Bonici}, \& {Bianchi}}]{Cagliari:2023mkq}
{Cagliari}, M.~S., {Castorina}, E., {Bonici}, M., \& {Bianchi}, D. 2024, \href{http://dx.doi.org/10.1088/1475-7516/2024/08/036}{\textcolor{blue}JCAP}, \href{https://ui.adsabs.harvard.edu/abs/2024JCAP...08..036C}{2024, 036}

\bibitem[{{Castorina} {et~al.}(2019){Castorina}, {Hand}, {Seljak}, {Beutler}, {Chuang}, {Zhao}, {Gil-Mar{\'\i}n}, {Percival}, {Ross}, {Choi}, {Dawson}, {de la Macorra}, {Rossi}, {Ruggeri}, {Schneider}, \& {Zhao}}]{Castorina2019}
{Castorina}, E., {Hand}, N., {Seljak}, U., {et~al.} 2019, \href{http://dx.doi.org/10.1088/1475-7516/2019/09/010}{\textcolor{blue}JCAP}, \href{https://ui.adsabs.harvard.edu/abs/2019JCAP...09..010C}{2019, 010}

\bibitem[{{Chaussidon} {et~al.}(2025){Chaussidon}, {Y{\`e}che}, {de Mattia}, {Payerne}, {McDonald}, {Ross}, {Ahlen}, {Bianchi}, {Brooks}, {Burtin}, {Claybaugh}, {de la Macorra}, {Doel}, {Ferraro}, {Font-Ribera}, {Forero-Romero}, {Gazta{\~n}aga}, {Gil-Mar{\'\i}n}, {Gontcho}, {Gutierrez}, {Guy}, {Honscheid}, {Howlett}, {Huterer}, {Kehoe}, {Kirkby}, {Kisner}, {Kremin}, {Le Guillou}, {Levi}, {Manera}, {Meisner}, {Miquel}, {Moustakas}, {Newman}, {Niz}, {Palanque-Delabrouille}, {Percival}, {Prada}, {P{\'e}rez-R{\`a}fols}, {Ravoux}, {Rossi}, {Sanchez}, {Schlegel}, {Schubnell}, {Seo}, {Sprayberry}, {Tarl{\'e}}, {Vargas-Maga{\~n}a}, {Weaver}, {Zhao}, \& {Zou}}]{DESI:2024oco}
{Chaussidon}, E., {Y{\`e}che}, C., {de Mattia}, A., {et~al.} 2025, \href{http://dx.doi.org/10.1088/1475-7516/2025/06/029}{\textcolor{blue}JCAP}, \href{https://ui.adsabs.harvard.edu/abs/2025JCAP...06..029C}{2025, 029}

\bibitem[{{Chen} {et~al.}(2025){Chen}, {Padmanabhan}, \& {Eisenstein}}]{Chen2025}
{Chen}, X., {Padmanabhan}, N., \& {Eisenstein}, D.~J. 2025, \href{http://dx.doi.org/10.1088/1475-7516/2025/08/055}{\textcolor{blue}JCAP}, \href{https://ui.adsabs.harvard.edu/abs/2025JCAP...08..055C}{2025, 055}

\bibitem[{{Coulton} {et~al.}(2024){Coulton}, {Abel}, \& {Banerjee}}]{Coulton:2023ouk}
{Coulton}, W.~R., {Abel}, T., \& {Banerjee}, A. 2024, \href{http://dx.doi.org/10.1093/mnras/stae2108}{\textcolor{blue}\mnras}, \href{https://ui.adsabs.harvard.edu/abs/2024MNRAS.534.1621C}{534, 1621}

\bibitem[{{Coulton} {et~al.}(2023{\natexlab{a}}){Coulton}, {Villaescusa-Navarro}, {Jamieson}, {Baldi}, {Jung}, {Karagiannis}, {Liguori}, {Verde}, \& {Wandelt}}]{Coulton:2022qbc}
{Coulton}, W.~R., {Villaescusa-Navarro}, F., {Jamieson}, D., {et~al.} 2023{\natexlab{a}}, \href{http://dx.doi.org/10.3847/1538-4357/aca8a7}{\textcolor{blue}\apj}, \href{https://ui.adsabs.harvard.edu/abs/2023ApJ...943...64C}{943, 64}

\bibitem[{{Coulton} {et~al.}(2023{\natexlab{b}}){Coulton}, {Villaescusa-Navarro}, {Jamieson}, {Baldi}, {Jung}, {Karagiannis}, {Liguori}, {Verde}, \& {Wandelt}}]{Coulton:2022rir}
{Coulton}, W.~R., {Villaescusa-Navarro}, F., {Jamieson}, D., {et~al.} 2023{\natexlab{b}}, \href{http://dx.doi.org/10.3847/1538-4357/aca7c1}{\textcolor{blue}\apj}, \href{https://ui.adsabs.harvard.edu/abs/2023ApJ...943..178C}{943, 178}

\bibitem[{{Creminelli} \& {Zaldarriaga}(2004)}]{Creminelli:2004yq}
{Creminelli}, P. \& {Zaldarriaga}, M. 2004, \href{http://dx.doi.org/10.1088/1475-7516/2004/10/006}{\textcolor{blue}JCAP}, \href{https://ui.adsabs.harvard.edu/abs/2004JCAP...10..006C}{2004, 006}

\bibitem[{{Crill} {et~al.}(2020){Crill}, {Werner}, {Akeson}, {Ashby}, {Bleem}, {Bock}, {Bryan}, {Burnham}, {Byunh}, {Chang}, {Chiang}, {Cook}, {Cooray}, {Davis}, {Dor{\'e}}, {Dowell}, {Dubois-Felsmann}, {Eifler}, {Faisst}, {Habib}, {Heinrich}, {Heitmann}, {Heaton}, {Hirata}, {Hristov}, {Hui}, {Jeong}, {Kang}, {Kecman}, {Kirkpatrick}, {Korngut}, {Krause}, {Lee}, {Lisse}, {Masters}, {Mauskopf}, {Melnick}, {Miyasaka}, {Nayyeri}, {Nguyen}, {{\"O}berg}, {Padin}, {Paladini}, {Pourrahmani}, {Pyo}, {Smith}, {Song}, {Symons}, {Teplitz}, {Tolls}, {Unwin}, {Windhorst}, {Yang}, \& {Zemcov}}]{spherex}
{Crill}, B.~P., {Werner}, M., {Akeson}, R., {et~al.} 2020, in Society of Photo-Optical Instrumentation Engineers (SPIE) Conference Series, Vol. 11443, Space Telescopes and Instrumentation 2020: Optical, Infrared, and Millimeter Wave, ed. M.~{Lystrup} \& M.~D. {Perrin}, \href{https://ui.adsabs.harvard.edu/abs/2020SPIE11443E..0IC}{114430I}

\bibitem[{{Dalal} {et~al.}(2008){Dalal}, {Dor{\'e}}, {Huterer}, \& {Shirokov}}]{Dalal2008}
{Dalal}, N., {Dor{\'e}}, O., {Huterer}, D., \& {Shirokov}, A. 2008, \href{http://dx.doi.org/10.1103/PhysRevD.77.123514}{\textcolor{blue}\prd}, \href{https://ui.adsabs.harvard.edu/abs/2008PhRvD..77l3514D}{77, 123514}

\bibitem[{{D'Amico} {et~al.}(2025){D'Amico}, {Lewandowski}, {Senatore}, \& {Zhang}}]{DAmico:2022gki}
{D'Amico}, G., {Lewandowski}, M., {Senatore}, L., \& {Zhang}, P. 2025, \href{http://dx.doi.org/10.1103/PhysRevD.111.063514}{\textcolor{blue}\prd}, \href{https://ui.adsabs.harvard.edu/abs/2025PhRvD.111f3514D}{111, 063514}

\bibitem[{{Davis} {et~al.}(1985){Davis}, {Efstathiou}, {Frenk}, \& {White}}]{FoF}
{Davis}, M., {Efstathiou}, G., {Frenk}, C.~S., \& {White}, S.~D.~M. 1985, \href{http://dx.doi.org/10.1086/163168}{\textcolor{blue}\apj}, \href{https://ui.adsabs.harvard.edu/abs/1985ApJ...292..371D}{292, 371}

\bibitem[{{DESI Collaboration} {et~al.}(2016){DESI Collaboration}, {Aghamousa}, {Aguilar}, {Ahlen}, {Alam}, {Allen}, {Allende Prieto}, {Annis}, {Bailey}, {Balland}, {Ballester}, {Baltay}, {Beaufore}, {Bebek}, {Beers}, {Bell}, {Bernal}, {Besuner}, {Beutler}, {Blake}, {Bleuler}, {Blomqvist}, {Blum}, {Bolton}, {Briceno}, {Brooks}, {Brownstein}, {Buckley-Geer}, {Burden}, {Burtin}, {Busca}, {Cahn}, {Cai}, {Cardiel-Sas}, {Carlberg}, {Carton}, {Casas}, {Castander}, {Cervantes-Cota}, {Claybaugh}, {Close}, {Coker}, {Cole}, {Comparat}, {Cooper}, {Cousinou}, {Crocce}, {Cuby}, {Cunningham}, {Davis}, {Dawson}, {de la Macorra}, {De Vicente}, {Delubac}, {Derwent}, {Dey}, {Dhungana}, {Ding}, {Doel}, {Duan}, {Ealet}, {Edelstein}, {Eftekharzadeh}, {Eisenstein}, {Elliott}, {Escoffier}, {Evatt}, {Fagrelius}, {Fan}, {Fanning}, {Farahi}, {Farihi}, {Favole}, {Feng}, {Fernandez}, {Findlay}, {Finkbeiner}, {Fitzpatrick}, {Flaugher}, {Flender}, {Font-Ribera}, {Forero-Romero}, {Fosalba}, {Frenk}, {Fumagalli}, {Gaensicke}, {Gallo},
  {Garcia-Bellido}, {Gaztanaga}, {Pietro Gentile Fusillo}, {Gerard}, {Gershkovich}, {Giannantonio}, {Gillet}, {Gonzalez-de-Rivera}, {Gonzalez-Perez}, {Gott}, {Graur}, {Gutierrez}, {Guy}, {Habib}, {Heetderks}, {Heetderks}, {Heitmann}, {Hellwing}, {Herrera}, {Ho}, {Holland}, {Honscheid}, {Huff}, {Hutchinson}, {Huterer}, {Hwang}, {Illa Laguna}, {Ishikawa}, {Jacobs}, {Jeffrey}, {Jelinsky}, {Jennings}, {Jiang}, {Jimenez}, {Johnson}, {Joyce}, {Jullo}, {Juneau}, {Kama}, {Karcher}, {Karkar}, {Kehoe}, {Kennamer}, {Kent}, {Kilbinger}, {Kim}, {Kirkby}, {Kisner}, {Kitanidis}, {Kneib}, {Koposov}, {Kovacs}, {Koyama}, {Kremin}, {Kron}, {Kronig}, {Kueter-Young}, {Lacey}, {Lafever}, {Lahav}, {Lambert}, {Lampton}, {Landriau}, {Lang}, {Lauer}, {Le Goff}, {Le Guillou}, {Le Van Suu}, {Lee}, {Lee}, {Leitner}, {Lesser}, {Levi}, {L'Huillier}, {Li}, {Liang}, {Lin}, {Linder}, {Loebman}, {Luki{\'c}}, {Ma}, {MacCrann}, {Magneville}, {Makarem}, {Manera}, {Manser}, {Marshall}, {Martini}, {Massey}, {Matheson}, {McCauley}, {McDonald},
  {McGreer}, {Meisner}, {Metcalfe}, {Miller}, {Miquel}, {Moustakas}, {Myers}, {Naik}, {Newman}, {Nichol}, {Nicola}, {Nicolati da Costa}, {Nie}, {Niz}, {Norberg}, {Nord}, {Norman}, {Nugent}, {O'Brien}, {Oh}, {Olsen}, {Padilla}, {Padmanabhan}, {Padmanabhan}, {Palanque-Delabrouille}, {Palmese}, {Pappalardo}, {P{\^a}ris}, {Park}, {Patej}, {Peacock}, {Peiris}, {Peng}, {Percival}, {Perruchot}, {Pieri}, {Pogge}, {Pollack}, {Poppett}, {Prada}, {Prakash}, {Probst}, {Rabinowitz}, {Raichoor}, {Ree}, {Refregier}, {Regal}, {Reid}, {Reil}, {Rezaie}, {Rockosi}, {Roe}, {Ronayette}, {Roodman}, {Ross}, {Ross}, {Rossi}, {Rozo}, {Ruhlmann-Kleider}, {Rykoff}, {Sabiu}, {Samushia}, {Sanchez}, {Sanchez}, {Schlegel}, {Schneider}, {Schubnell}, {Secroun}, {Seljak}, {Seo}, {Serrano}, {Shafieloo}, {Shan}, {Sharples}, {Sholl}, {Shourt}, {Silber}, {Silva}, {Sirk}, {Slosar}, {Smith}, {Smoot}, {Som}, {Song}, {Sprayberry}, {Staten}, {Stefanik}, {Tarle}, {Sien Tie}, {Tinker}, {Tojeiro}, {Valdes}, {Valenzuela}, {Valluri}, {Vargas-Magana},
  {Verde}, {Walker}, {Wang}, {Wang}, {Weaver}, {Weaverdyck}, {Wechsler}, {Weinberg}, {White}, {Yang}, {Yeche}, {Zhang}, {Zhao}, {Zheng}, {Zhou}, {Zhou}, {Zhu}, {Zou}, \& {Zu}}]{desi}
{DESI Collaboration}, {Aghamousa}, A., {Aguilar}, J., {et~al.} 2016, \href{https://ui.adsabs.harvard.edu/abs/2016arXiv161100036D}{\href{http://dx.doi.org/10.48550/arXiv.1611.00036}{\textcolor{blue}arXiv e-prints}, arXiv:1611.00036}

\bibitem[{{Desjacques} \& {Seljak}(2010)}]{Desjaques2010}
{Desjacques}, V. \& {Seljak}, U. 2010, \href{http://dx.doi.org/10.1155/2010/908640}{\textcolor{blue}Advances in Astronomy}, \href{https://ui.adsabs.harvard.edu/abs/2010AdAst2010E..89D}{2010, 908640}

\bibitem[{{Euclid Collaboration: {Mellier}} {et~al.}(2025){Euclid Collaboration: {Mellier}}, {Abdurro'uf}, {Acevedo Barroso}, {Ach{\'u}carro}, {Adamek}, {Adam}, {Addison}, {Aghanim}, {Aguena}, {Ajani}, {Akrami}, {Al-Bahlawan}, {Alavi}, {Albuquerque}, {Alestas}, {Alguero}, {Allaoui}, {Allen}, {Allevato}, {Alonso-Tetilla}, {Altieri}, {Alvarez-Candal}, {Alvi}, {Amara}, {Amendola}, {Amiaux}, {Andika}, {Andreon}, {Andrews}, {Angora}, {Angulo}, {Annibali}, {Anselmi}, {Anselmi}, {Arcari}, {Archidiacono}, {Aric{\`o}}, {Arnaud}, {Arnouts}, {Asgari}, {Asorey}, {Atayde}, {Atek}, {Atrio-Barandela}, {Aubert}, {Aubourg}, {Auphan}, {Auricchio}, {Aussel}, {Aussel}, {Avelino}, {Avgoustidis}, {Avila}, {Awan}, {Azzollini}, {Baccigalupi}, {Bachelet}, {Bacon}, {Baes}, {Bagley}, {Bahr-Kalus}, {Balaguera-Antolinez}, {Balbinot}, {Balcells}, {Baldi}, {Baldry}, {Balestra}, {Ballardini}, {Ballester}, {Balogh}, {Ba{\~n}ados}, {Barbier}, {Bardelli}, {Baron}, {Barreiro}, {Barrena}, {Barriere}, {Barros}, {Barthelemy}, {Bartolo}, {Basset},
  {Battaglia}, {Battisti}, {Baugh}, {Baumont}, {Bazzanini}, {Beaulieu}, {Beckmann}, {Belikov}, {Bel}, {Bellagamba}, {Bella}, {Bellini}, {Benabed}, {Bender}, {Benevento}, {Bennett}, {Benson}, {Bergamini}, {Bermejo-Climent}, {Bernardeau}, {Bertacca}, {Berthe}, {Berthier}, {Bethermin}, {Beutler}, {Bevillon}, {Bhargava}, {Bhatawdekar}, {Bianchi}, {Bisigello}, {Biviano}, {Blake}, {Blanchard}, {Blazek}, {Blot}, {Bosco}, {Bodendorf}, {Boenke}, {B{\"o}hringer}, {Boldrini}, {Bolzonella}, {Bonchi}, {Bonici}, {Bonino}, {Bonino}, {Bonvin}, {Bon}, {Booth}, {Borgani}, {Borlaff}, {Borsato}, {Bose}, {Botticella}, {Boucaud}, {Bouche}, {Boucher}, {Boutigny}, {Bouvard}, {Bouwens}, {Bouy}, {Bowler}, {Bozza}, {Bozzo}, {Branchini}, {Brando}, {Brau-Nogue}, {Brekke}, {Bremer}, {Brescia}, {Breton}, {Brinchmann}, {Brinckmann}, {Brockley-Blatt}, {Brodwin}, {Brouard}, {Brown}, {Bruton}, {Bucko}, {Buddelmeijer}, {Buenadicha}, {Buitrago}, {Burger}, {Burigana}, {Busillo}, {Busonero}, {Cabanac}, {Cabayol-Garcia}, {Cagliari}, {Caillat},
  {Caillat}, {Calabrese}, {Calabro}, {Calderone}, {Calura}, {Camacho Quevedo}, {Camera}, {Campos}, {Ca{\~n}as-Herrera}, {Candini}, {Cantiello}, {Capobianco}, {Cappellaro}, {Cappelluti}, {Cappi}, {Caputi}, {Cara}, {Carbone}, {Cardone}, {Carella}, {Carlberg}, {Carle}, {Carminati}, {Caro}, {Carrasco}, {Carretero}, {Carrilho}, {Carron Duque}, \& {Carry}}]{Euclid:2024yrr}
{Euclid Collaboration: {Mellier}}, Y., {Abdurro'uf}, {Acevedo Barroso}, J.~A., {et~al.} 2025, \href{http://dx.doi.org/10.1051/0004-6361/202450810}{\textcolor{blue}\aap}, \href{https://ui.adsabs.harvard.edu/abs/2025A&A...697A...1E}{697, A1}

\bibitem[{{Euclid Collaboration: {Scaramella}} {et~al.}(2022){Euclid Collaboration: {Scaramella}}, {Amiaux}, {Mellier}, {Burigana}, {Carvalho}, {Cuillandre}, {Da Silva}, {Derosa}, {Dinis}, {Maiorano}, {Maris}, {Tereno}, {Laureijs}, {Boenke}, {Buenadicha}, {Dupac}, {Gaspar Venancio}, {G{\'o}mez-{\'A}lvarez}, {Hoar}, {Lorenzo Alvarez}, {Racca}, {Saavedra-Criado}, {Schwartz}, {Vavrek}, {Schirmer}, {Aussel}, {Azzollini}, {Cardone}, {Cropper}, {Ealet}, {Garilli}, {Gillard}, {Granett}, {Guzzo}, {Hoekstra}, {Jahnke}, {Kitching}, {Maciaszek}, {Meneghetti}, {Miller}, {Nakajima}, {Niemi}, {Pasian}, {Percival}, {Pottinger}, {Sauvage}, {Scodeggio}, {Wachter}, {Zacchei}, {Aghanim}, {Amara}, {Auphan}, {Auricchio}, {Awan}, {Balestra}, {Bender}, {Bodendorf}, {Bonino}, {Branchini}, {Brau-Nogue}, {Brescia}, {Candini}, {Capobianco}, {Carbone}, {Carlberg}, {Carretero}, {Casas}, {Castander}, {Castellano}, {Cavuoti}, {Cimatti}, {Cledassou}, {Congedo}, {Conselice}, {Conversi}, {Copin}, {Corcione}, {Costille}, {Courbin},
  {Degaudenzi}, {Douspis}, {Dubath}, {Duncan}, {Dusini}, {Farrens}, {Ferriol}, {Fosalba}, {Fourmanoit}, {Frailis}, {Franceschi}, {Franzetti}, {Fumana}, {Gillis}, {Giocoli}, {Grazian}, {Grupp}, {Haugan}, {Holmes}, {Hormuth}, {Hudelot}, {Kermiche}, {Kiessling}, {Kilbinger}, {Kohley}, {Kubik}, {K{\"u}mmel}, {Kunz}, {Kurki-Suonio}, {Lahav}, {Ligori}, {Lilje}, {Lloro}, {Mansutti}, {Marggraf}, {Markovic}, {Marulli}, {Massey}, {Maurogordato}, {Melchior}, {Merlin}, {Meylan}, {Mohr}, {Moresco}, {Morin}, {Moscardini}, {Munari}, {Nichol}, {Padilla}, {Paltani}, {Peacock}, {Pedersen}, {Pettorino}, {Pires}, {Poncet}, {Popa}, {Pozzetti}, {Raison}, {Rebolo}, {Rhodes}, {Rix}, {Roncarelli}, {Rossetti}, {Saglia}, {Schneider}, {Schrabback}, {Secroun}, {Seidel}, {Serrano}, {Sirignano}, {Sirri}, {Skottfelt}, {Stanco}, {Starck}, {Tallada-Cresp{\'\i}}, {Tavagnacco}, {Taylor}, {Teplitz}, {Toledo-Moreo}, {Torradeflot}, {Trifoglio}, {Valentijn}, {Valenziano}, {Verdoes Kleijn}, {Wang}, {Welikala}, {Weller}, {Wetzstein}, {Zamorani},
  {Zoubian}, {Andreon}, {Baldi}, {Bardelli}, {Boucaud}, {Camera}, {Di Ferdinando}, {Fabbian}, {Farinelli}, {Galeotta}, {Graci{\'a}-Carpio}, {Maino}, {Medinaceli}, {Mei}, {Neissner}, {Polenta}, {Renzi}, {Romelli}, {Rosset}, {Sureau}, {Tenti}, {Vassallo}, {Zucca}, {Baccigalupi}, {Balaguera-Antol{\'\i}nez}, {Battaglia}, {Biviano}, {Borgani}, {Bozzo}, {Cabanac}, {Cappi}, {Casas}, {Castignani}, {Colodro-Conde}, {Coupon}, {Courtois}, {Cuby}, {de la Torre}, {Desai}, {Dole}, {Fabricius}, {Farina}, {Ferreira}, {Finelli}, {Flose-Reimberg}, {Fotopoulou}, {Ganga}, {Gozaliasl}, {Hook}, {Keihanen}, {Kirkpatrick}, {Liebing}, {Lindholm}, {Mainetti}, {Martinelli}, {Martinet}, {Maturi}, {McCracken}, {Metcalf}, {Morgante}, {Nightingale}, {Nucita}, {Patrizii}, {Potter}, {Riccio}, {S{\'a}nchez}, {Sapone}, {Schewtschenko}, {Schultheis}, {Scottez}, {Teyssier}, {Tutusaus}, {Valiviita}, {Viel}, {Vriend}, \& {Whittaker}}]{EuclidWide}
{Euclid Collaboration: {Scaramella}}, R., {Amiaux}, J., {Mellier}, Y., {et~al.} 2022, \href{http://dx.doi.org/10.1051/0004-6361/202141938}{\textcolor{blue}\aap}, \href{https://ui.adsabs.harvard.edu/abs/2022A&A...662A.112E}{662, A112}

\bibitem[{{Giri} {et~al.}(2023){Giri}, {M{\"u}nchmeyer}, \& {Smith}}]{Giri2023}
{Giri}, U., {M{\"u}nchmeyer}, M., \& {Smith}, K.~M. 2023, \href{http://dx.doi.org/10.1103/PhysRevD.107.L061301}{\textcolor{blue}\prd}, \href{https://ui.adsabs.harvard.edu/abs/2023PhRvD.107f1301G}{107, L061301}

\bibitem[{{Hahn} {et~al.}(2020){Hahn}, {Villaescusa-Navarro}, {Castorina}, \& {Scoccimarro}}]{2020JCAP...03..040H}
{Hahn}, C., {Villaescusa-Navarro}, F., {Castorina}, E., \& {Scoccimarro}, R. 2020, \href{http://dx.doi.org/10.1088/1475-7516/2020/03/040}{\textcolor{blue}JCAP}, \href{https://ui.adsabs.harvard.edu/abs/2020JCAP...03..040H}{2020, 040}

\bibitem[{{Hartlap} {et~al.}(2007){Hartlap}, {Simon}, \& {Schneider}}]{Hartlap-factor}
{Hartlap}, J., {Simon}, P., \& {Schneider}, P. 2007, \href{http://dx.doi.org/10.1051/0004-6361:20066170}{\textcolor{blue}\aap}, \href{https://ui.adsabs.harvard.edu/abs/2007A&A...464..399H}{464, 399}

\bibitem[{{Jung} {et~al.}(2023{\natexlab{a}}){Jung}, {Karagiannis}, {Liguori}, {Baldi}, {Coulton}, {Jamieson}, {Verde}, {Villaescusa-Navarro}, \& {Wandelt}}]{Jung:2022gfa}
{Jung}, G., {Karagiannis}, D., {Liguori}, M., {et~al.} 2023{\natexlab{a}}, \href{http://dx.doi.org/10.3847/1538-4357/acc4bd}{\textcolor{blue}\apj}, \href{https://ui.adsabs.harvard.edu/abs/2023ApJ...948..135J}{948, 135}

\bibitem[{{Jung} {et~al.}(2023{\natexlab{b}}){Jung}, {Ravenni}, {Baldi}, {Coulton}, {Jamieson}, {Karagiannis}, {Liguori}, {Shao}, {Verde}, {Villaescusa-Navarro}, \& {Wandelt}}]{Jung:2023kjh}
{Jung}, G., {Ravenni}, A., {Baldi}, M., {et~al.} 2023{\natexlab{b}}, \href{http://dx.doi.org/10.3847/1538-4357/acfe70}{\textcolor{blue}\apj}, \href{https://ui.adsabs.harvard.edu/abs/2023ApJ...957...50J}{957, 50}

\bibitem[{{Jung} {et~al.}(2024){Jung}, {Ravenni}, {Liguori}, {Baldi}, {Coulton}, {Villaescusa-Navarro}, \& {Wandelt}}]{Jung:2024esv}
{Jung}, G., {Ravenni}, A., {Liguori}, M., {et~al.} 2024, \href{http://dx.doi.org/10.3847/1538-4357/ad83bd}{\textcolor{blue}\apj}, \href{https://ui.adsabs.harvard.edu/abs/2024ApJ...976..109J}{976, 109}

\bibitem[{{Kvasiuk} {et~al.}(2025){Kvasiuk}, {M{\"u}nchmeyer}, \& {Smith}}]{Kvasiuk:2024gbz}
{Kvasiuk}, Y., {M{\"u}nchmeyer}, M., \& {Smith}, K. 2025, \href{http://dx.doi.org/10.1103/2szy-wypg}{\textcolor{blue}\prd}, \href{https://ui.adsabs.harvard.edu/abs/2025PhRvD.112b3540K}{112, 023540}

\bibitem[{LeCun(1989)}]{LeCun:89}
LeCun, Y. 1989, Generalization and network design strategies, ed. R.~Pfeifer, Z.~Schreter, F.~Fogelman, \& L.~Steels (Elsevier)

\bibitem[{{LSST Science Collaboration} {et~al.}(2009){LSST Science Collaboration}, {Abell}, {Allison}, {Anderson}, {Andrew}, {Angel}, {Armus}, {Arnett}, {Asztalos}, {Axelrod}, {Bailey}, {Ballantyne}, {Bankert}, {Barkhouse}, {Barr}, {Barrientos}, {Barth}, {Bartlett}, {Becker}, {Becla}, {Beers}, {Bernstein}, {Biswas}, {Blanton}, {Bloom}, {Bochanski}, {Boeshaar}, {Borne}, {Bradac}, {Brandt}, {Bridge}, {Brown}, {Brunner}, {Bullock}, {Burgasser}, {Burge}, {Burke}, {Cargile}, {Chandrasekharan}, {Chartas}, {Chesley}, {Chu}, {Cinabro}, {Claire}, {Claver}, {Clowe}, {Connolly}, {Cook}, {Cooke}, {Cooray}, {Covey}, {Culliton}, {de Jong}, {de Vries}, {Debattista}, {Delgado}, {Dell'Antonio}, {Dhital}, {Di Stefano}, {Dickinson}, {Dilday}, {Djorgovski}, {Dobler}, {Donalek}, {Dubois-Felsmann}, {Durech}, {Eliasdottir}, {Eracleous}, {Eyer}, {Falco}, {Fan}, {Fassnacht}, {Ferguson}, {Fernandez}, {Fields}, {Finkbeiner}, {Figueroa}, {Fox}, {Francke}, {Frank}, {Frieman}, {Fromenteau}, {Furqan}, {Galaz}, {Gal-Yam}, {Garnavich},
  {Gawiser}, {Geary}, {Gee}, {Gibson}, {Gilmore}, {Grace}, {Green}, {Gressler}, {Grillmair}, {Habib}, {Haggerty}, {Hamuy}, {Harris}, {Hawley}, {Heavens}, {Hebb}, {Henry}, {Hileman}, {Hilton}, {Hoadley}, {Holberg}, {Holman}, {Howell}, {Infante}, {Ivezic}, {Jacoby}, {Jain}, {R}, {Jedicke}, {Jee}, {Garrett Jernigan}, {Jha}, {Johnston}, {Jones}, {Juric}, {Kaasalainen}, {Styliani}, {Kafka}, {Kahn}, {Kaib}, {Kalirai}, {Kantor}, {Kasliwal}, {Keeton}, {Kessler}, {Knezevic}, {Kowalski}, {Krabbendam}, {Krughoff}, {Kulkarni}, {Kuhlman}, {Lacy}, {Lepine}, {Liang}, {Lien}, {Lira}, {Long}, {Lorenz}, {Lotz}, {Lupton}, {Lutz}, {Macri}, {Mahabal}, {Mandelbaum}, {Marshall}, {May}, {McGehee}, {Meadows}, {Meert}, {Milani}, {Miller}, {Miller}, {Mills}, {Minniti}, {Monet}, {Mukadam}, {Nakar}, {Neill}, {Newman}, {Nikolaev}, {Nordby}, {O'Connor}, {Oguri}, {Oliver}, {Olivier}, {Olsen}, {Olsen}, {Olszewski}, {Oluseyi}, {Padilla}, {Parker}, {Pepper}, {Peterson}, {Petry}, {Pinto}, {Pizagno}, {Popescu}, {Prsa}, {Radcka}, {Raddick},
  {Rasmussen}, {Rau}, {Rho}, {Rhoads}, {Richards}, {Ridgway}, {Robertson}, {Roskar}, {Saha}, {Sarajedini}, {Scannapieco}, {Schalk}, {Schindler}, {Schmidt}, {Schmidt}, {Schneider}, {Schumacher}, {Scranton}, {Sebag}, {Seppala}, {Shemmer}, {Simon}, {Sivertz}, {Smith}, {Allyn Smith}, {Smith}, {Spitz}, {Stanford}, {Stassun}, {Strader}, {Strauss}, {Stubbs}, {Sweeney}, {Szalay}, {Szkody}, {Takada}, {Thorman}, {Trilling}, {Trimble}, {Tyson}, {Van Berg}, {Vanden Berk}, {VanderPlas}, {Verde}, {Vrsnak}, {Walkowicz}, {Wandelt}, {Wang}, {Wang}, {Warner}, {Wechsler}, {West}, {Wiecha}, {Williams}, {Willman}, {Wittman}, {Wolff}, {Wood-Vasey}, {Wozniak}, {Young}, {Zentner}, \& {Zhan}}]{lsst}
{LSST Science Collaboration}, {Abell}, P.~A., {Allison}, J., {et~al.} 2009, \href{https://ui.adsabs.harvard.edu/abs/2009arXiv0912.0201L}{\href{http://dx.doi.org/10.48550/arXiv.0912.0201}{\textcolor{blue}arXiv e-prints}, arXiv:0912.0201}

\bibitem[{Maas(2013)}]{Maas2013RectifierNI}
Maas, A.~L. 2013, \href{}{in }

\bibitem[{{Maldacena}(2003)}]{Maldacena:2002vr}
{Maldacena}, J. 2003, \href{http://dx.doi.org/10.1088/1126-6708/2003/05/013}{\textcolor{blue}Journal of High Energy Physics}, \href{https://ui.adsabs.harvard.edu/abs/2003JHEP...05..013M}{2003, 013}

\bibitem[{{Marinucci} {et~al.}(2025){Marinucci}, {Jung}, {Liguori}, {Ravenni}, {Spezzati}, {Andrews}, {Baldi}, {Coulton}, {Karagiannis}, {Villaescusa-Navarro}, \& {Wandelt}}]{Marinucci2025}
{Marinucci}, M., {Jung}, G., {Liguori}, M., {et~al.} 2025, \href{http://dx.doi.org/10.1088/1475-7516/2025/09/036}{\textcolor{blue}JCAP}, \href{https://ui.adsabs.harvard.edu/abs/2025JCAP...09..036M}{2025, 036}

\bibitem[{{Nagarajappa} \& {Ma}(2024)}]{Nagarajappa2024}
{Nagarajappa}, C.~G. \& {Ma}, Y.-Z. 2024, \href{http://dx.doi.org/10.1093/mnras/stae679}{\textcolor{blue}\mnras}, \href{https://ui.adsabs.harvard.edu/abs/2024MNRAS.529.3289N}{529, 3289}

\bibitem[{{Peron} {et~al.}(2024){Peron}, {Jung}, {Liguori}, \& {Pietroni}}]{Peron2024}
{Peron}, M., {Jung}, G., {Liguori}, M., \& {Pietroni}, M. 2024, \href{http://dx.doi.org/10.1088/1475-7516/2024/07/021}{\textcolor{blue}JCAP}, \href{https://ui.adsabs.harvard.edu/abs/2024JCAP...07..021P}{2024, 021}

\bibitem[{{Planck Collaboration} {et~al.}(2020){Planck Collaboration}, {Akrami}, {Arroja}, {Ashdown}, {Aumont}, {Baccigalupi}, {Ballardini}, {Banday}, {Barreiro}, {Bartolo}, {Basak}, {Benabed}, {Bernard}, {Bersanelli}, {Bielewicz}, {Bond}, {Borrill}, {Bouchet}, {Bucher}, {Burigana}, {Butler}, {Calabrese}, {Cardoso}, {Casaponsa}, {Challinor}, {Chiang}, {Colombo}, {Combet}, {Crill}, {Cuttaia}, {de Bernardis}, {de Rosa}, {de Zotti}, {Delabrouille}, {Delouis}, {Di Valentino}, {Diego}, {Dor{\'e}}, {Douspis}, {Ducout}, {Dupac}, {Dusini}, {Efstathiou}, {Elsner}, {En{\ss}lin}, {Eriksen}, {Fantaye}, {Fergusson}, {Fernandez-Cobos}, {Finelli}, {Frailis}, {Fraisse}, {Franceschi}, {Frolov}, {Galeotta}, {Galli}, {Ganga}, {G{\'e}nova-Santos}, {Gerbino}, {Gonz{\'a}lez-Nuevo}, {G{\'o}rski}, {Gratton}, {Gruppuso}, {Gudmundsson}, {Hamann}, {Handley}, {Hansen}, {Herranz}, {Hivon}, {Huang}, {Jaffe}, {Jones}, {Jung}, {Keih{\"a}nen}, {Keskitalo}, {Kiiveri}, {Kim}, {Krachmalnicoff}, {Kunz}, {Kurki-Suonio}, {Lamarre}, {Lasenby},
  {Lattanzi}, {Lawrence}, {Le Jeune}, {Levrier}, {Lewis}, {Liguori}, {Lilje}, {Lindholm}, {L{\'o}pez-Caniego}, {Ma}, {Mac{\'\i}as-P{\'e}rez}, {Maggio}, {Maino}, {Mandolesi}, {Marcos-Caballero}, {Maris}, {Martin}, {Mart{\'\i}nez-Gonz{\'a}lez}, {Matarrese}, {Mauri}, {McEwen}, {Meerburg}, {Meinhold}, {Melchiorri}, {Mennella}, {Migliaccio}, {Miville-Desch{\^e}nes}, {Molinari}, {Moneti}, {Montier}, {Morgante}, {Moss}, {M{\"u}nchmeyer}, {Natoli}, {Oppizzi}, {Pagano}, {Paoletti}, {Partridge}, {Patanchon}, {Perrotta}, {Pettorino}, {Piacentini}, {Polenta}, {Puget}, {Rachen}, {Racine}, {Reinecke}, {Remazeilles}, {Renzi}, {Rocha}, {Rubi{\~n}o-Mart{\'\i}n}, {Ruiz-Granados}, {Salvati}, {Savelainen}, {Scott}, {Shellard}, {Shiraishi}, {Sirignano}, {Sirri}, {Smith}, {Spencer}, {Stanco}, {Sunyaev}, {Suur-Uski}, {Tauber}, {Tavagnacco}, {Tenti}, {Toffolatti}, {Tomasi}, {Trombetti}, {Valiviita}, {Van Tent}, {Vielva}, {Villa}, {Vittorio}, {Wandelt}, {Wehus}, {Zacchei}, \& {Zonca}}]{2020A&A...641A...9P}
{Planck Collaboration}, {Akrami}, Y., {Arroja}, F., {et~al.} 2020, \href{http://dx.doi.org/10.1051/0004-6361/201935891}{\textcolor{blue}\aap}, \href{https://ui.adsabs.harvard.edu/abs/2020A&A...641A...9P}{641, A9}

\bibitem[{{Scoccimarro}(2015)}]{2015PhRvD..92h3532S}
{Scoccimarro}, R. 2015, \href{http://dx.doi.org/10.1103/PhysRevD.92.083532}{\textcolor{blue}\prd}, \href{https://ui.adsabs.harvard.edu/abs/2015PhRvD..92h3532S}{92, 083532}

\bibitem[{{Senatore} \& {Zaldarriaga}(2012)}]{Senatore:2010wk}
{Senatore}, L. \& {Zaldarriaga}, M. 2012, \href{http://dx.doi.org/10.1007/JHEP04(2012)024}{\textcolor{blue}Journal of High Energy Physics}, \href{https://ui.adsabs.harvard.edu/abs/2012JHEP...04..024S}{2012, 24}

\bibitem[{{Slosar} {et~al.}(2008){Slosar}, {Hirata}, {Seljak}, {Ho}, \& {Padmanabhan}}]{Slosar2008}
{Slosar}, A., {Hirata}, C., {Seljak}, U., {Ho}, S., \& {Padmanabhan}, N. 2008, \href{http://dx.doi.org/10.1088/1475-7516/2008/08/031}{\textcolor{blue}JCAP}, \href{https://ui.adsabs.harvard.edu/abs/2008JCAP...08..031S}{2008, 031}

\bibitem[{{Springel} {et~al.}(2008){Springel}, {Wang}, {Vogelsberger}, {Ludlow}, {Jenkins}, {Helmi}, {Navarro}, {Frenk}, \& {White}}]{2008MNRAS.391.1685S}
{Springel}, V., {Wang}, J., {Vogelsberger}, M., {et~al.} 2008, \href{http://dx.doi.org/10.1111/j.1365-2966.2008.14066.x}{\textcolor{blue}\mnras}, \href{https://ui.adsabs.harvard.edu/abs/2008MNRAS.391.1685S}{391, 1685}

\bibitem[{{Springel} {et~al.}(2005){Springel}, {White}, {Jenkins}, {Frenk}, {Yoshida}, {Gao}, {Navarro}, {Thacker}, {Croton}, {Helly}, {Peacock}, {Cole}, {Thomas}, {Couchman}, {Evrard}, {Colberg}, \& {Pearce}}]{gadget2}
{Springel}, V., {White}, S. D.~M., {Jenkins}, A., {et~al.} 2005, \href{http://dx.doi.org/10.1038/nature03597}{\textcolor{blue}\nat}, \href{https://ui.adsabs.harvard.edu/abs/2005Natur.435..629S}{435, 629}

\bibitem[{{Villaescusa-Navarro}(2018)}]{Pylians}
{Villaescusa-Navarro}, F. 2018, {Pylians: Python libraries for the analysis of numerical simulations}, Astrophysics Source Code Library, record ascl:1811.008

\bibitem[{{Villaescusa-Navarro} {et~al.}(2020){Villaescusa-Navarro}, {Hahn}, {Massara}, {Banerjee}, {Delgado}, {Ramanah}, {Charnock}, {Giusarma}, {Li}, {Allys}, {Brochard}, {Uhlemann}, {Chiang}, {He}, {Pisani}, {Obuljen}, {Feng}, {Castorina}, {Contardo}, {Kreisch}, {Nicola}, {Alsing}, {Scoccimarro}, {Verde}, {Viel}, {Ho}, {Mallat}, {Wandelt}, \& {Spergel}}]{Villaescusa-Navarro:2019bje}
{Villaescusa-Navarro}, F., {Hahn}, C., {Massara}, E., {et~al.} 2020, \href{http://dx.doi.org/10.3847/1538-4365/ab9d82}{\textcolor{blue}\apjs}, \href{https://ui.adsabs.harvard.edu/abs/2020ApJS..250....2V}{250, 2}

\end{thebibliography}

\end{document}